**Using AI in engineering education: a balancing act, driven by clear purpose**

**Olya Kudina**

o.kudina@tudelft.nl

TU Delft, The Netherlands

Optentia Unit, North-West University, South Africa

Draft version, to appear in *The Routledge Handbook of the Philosophy of Engineering*, 2nd ed.
Edited By Diane P. Michelfelder, Neelke Doorn

## Abstract

Based on a questionnaire of 100 higher-education students, predominantly from engineering-related fields, and a critical review of recent literature, this chapter examines how students use and perceive Large Language Models (LLMs) in engineering education. Students primarily value LLMs for writing support, conceptual clarification, coding assistance, and brainstorming, while simultaneously expressing concerns about inaccuracies, bias, overreliance, academic integrity, and the burden of verification. Through an analysis of two dominant metaphors, namely LLMs as an “oracle” and as a “tutor,” the chapter shows how these systems cultivate expectations of authority, expertise, and personalized learning that often exceed their actual capabilities as probabilistic text generators. The chapter further argues that students’ attachment to the promises of efficiency and personalized support reflects a form of “cruel optimism,” where the perceived benefits of LLMs often depend on the very skills, vigilance, and expertise that students are still developing. Overall, the chapter argues for a purpose-driven and context-sensitive approach to AI integration in engineering education, emphasizing critical AI literacy, reflective assessment design, pedagogical caution, and consideration of broader ethical and environmental impacts.

## Introduction

Artificial Intelligence (AI), especially Generative AI (GenAI), probabilistic systems that synthesize human-resembling text, image, and audio content based on the trained data, has rapidly penetrated society due to its ease of use and promises of multidimensional utility. It has also confronted engineering educators and students with finding ways to reap the potential benefits of AI’s use in the classroom while accounting for its ethical and practical considerations. Like any technology, AI systems mediate the way people think and how they act (Verbeek, 2005), affecting the values underlying these perceptions and decisions (Kudina, 2024). In this chapter, I will focus on the use of Large Language Models (LLMs) as the text-generative pillar of GenAI that attracted significant attention in engineering education through its user-friendly applications, such as ChatGPT, Co-pilot, and NotebookLM.

In education overall and in engineering education in particular, LLMs are associated with the promise of new and diverse ways of accessing knowledge (Giannakos et al., 2025), lowering access barriers regardless of the students' native language and promoting cultural diversity (Oyetade & Zuva, 2025), and, in particular, helping to formulate thoughts quickly and efficiently (Vartiainen et al., 2025). In parallel, the ethical worries about their use in class concern how LLMs change the value of education and knowledge production (Peterson, 2025), and the related values of creativity and research integrity (Arar et al., 2025; Köbis et al., 2025). The more concrete concerns frequently relate to LLMs reducing evaluation in information processing (Bender et al., 2021; Musi & Palmieri, 2024) and the overreliance of students on technological tools for writing (Fan et al., 2025), that, as some suggest (Kosmyna et al., 2025), gradually leads to cognitive debt.

For educators, the issue of potential use or non-use of AI became polarizing. Some advocate for its adoption, citing the need to teach students a new way to think with GenAI (Rogers, 2025), the responsibility to train students who can compete in the job market (Owoc et al., 2019), and the opportunity GenAI represents to optimize teaching practices (Granić, 2025). Additionally, the proponents of AI integration in learning cite that failure to do so would leave behind both the students (Giray, 2024; Hanshaw & Sullivan, 2025) and the educators (Chen & Xu, 2025), potentially resulting in educators' inequality: "Teachers are confronted with the choice of embracing new Artificial Intelligence (AI) technologies or risking obsolescence" (Memon & Kwan, 2025, p. 1). Others strongly oppose AI's introduction in higher education (e.g., Flenady & Sparrow, 2025; Guest et al., 2025), questioning the pedagogical utility and the personalization potential of LLMs (Laak et al., 2024), suggesting they decrease the learning skills in students (Bastani et al., 2024; Wieczorek, 2025), or contesting AI on principled bases (e.g., the role of big tech in education, sustainability concerns, pushing back on the increasing workload, etc.) (Bender et al., 2025). Both sides of the debate see the introduction of GenAI in education as a high-stakes situation and reflect the struggles to reason with it, because while some background information is already available, the technologically induced disruption is still underway, making it difficult to identify a decisive orientation.

As is typical for navigating such morally uncertain situations (Nickel et al., 2022), apart from polarization, the rapid introduction of LLMs has exposed many systemic issues in engineering education, including maximizing efficiency while balancing pedagogical value, balancing pedagogical innovation with the increasing workload and decreasing education budget (Wanyonyi & Murithi, 2025), maintaining educational independence while being increasingly reliant on digital tools and becoming locked into large corporate technological infrastructures without the ability to contest or question these systems (Abbas et al., 2024; Bender et al., 2025; Monett & Paquet, 2025), etc. It is no surprise that, several years after

LLMs became widely introduced, universities still struggle to identify best practices for their use.

Meanwhile, engineering students continue to experiment with and adopt this technology. Providing any kind of support necessitates, first and foremost, an understanding of *why students turn to LLMs in education*. With this chapter, I aim to offer a first-person perspective from students on the educational value they perceive in these systems, based on an analysis of a global educational questionnaire I conducted in 2023-24. This will allow me to outline and reflect on the typical use cases of LLMs in engineering education, accompanied by the students' impressions and literature analysis. Additionally, I will outline the typical metaphors that students identify LLMs' utility with (e.g., an oracle, a calculator) and reflect on them ethically. Finally, I will provide reflections on ways to move forward regarding LLMs in engineering education, accompanied by concrete starting questions, examples of reflective use of LLMs in engineering education, and other pointers. Overall, this chapter will provide a critical and empirically informed account of the use of LLMs in the classroom, suggesting a pragmatic approach to determining their use on a case-by-case basis and avoiding their use when a benefit is not clear or is outweighed by the ethical and environmental concerns.

**Why students turn to LLMs: a questionnaire analysis**

In this section, I will briefly describe the results of the international questionnaire on the use of LLMs in education that I conducted online in 2023-24. I wanted to conduct this questionnaire because I was increasingly confronted with students' use of LLMs in my diverse classes, which include teaching engineering students at TU Delft, bioethicists at Yale University, and technoanthropologists at Aalborg University. I wanted to understand why students turn to this technology, whether I should implement it in education and if so, how, and whether I could formulate some insights for educators based on these students' experiences. To be sure, I don't pretend to generalize from the questionnaire alone, but will accompany my conceptualizations with state-of-the-art literature studies and systematic reviews. Together, this will enable me to paint a more comprehensive picture and account for the size and timespan limitations of the questionnaire.

The questionnaire began by asking whether the respondents used LLMs for their studies, if yes, for which purpose and how often, asking the respondents next to provide their overall impressions, to specify the good and the problematic aspects of using LLMs, their opinion on the role of LLMs in studies, and the actions universities should take concerning LLMs. The quotes preserve the original spelling, adding [*sic*] after spelling mistakes. To ensure anonymous treatment of responses, the questionnaire did not collect personally identifiable information[1], only asking for the students' age range, level of higher education study, and the field of study. Most of the respondents were 18-24 y.o. (57%), with 25-34 y.o. (32%) a close second age group and only 11% older (35-64 y.o.). 59% reported pursuing graduate/master's level studies, 24% - undergraduate/bachelor's, and 17% - postgraduate/PhD level. The range of study fields was broad (see Table 1 below), from the social sciences and humanities to law and technology studies, but was dominated by the engineering field (Artificial Intelligence, Robotics, Engineering and Design disciplines represented 52% of the respondents), making the findings especially relevant for engineering education. 19 respondents have never used LLMs in their studies. Out of 81 that have, 22 students reported using LLMs for education daily, 27 - several times per week (2–4 days/week), 14 - weekly (about once a week), 13 reported using them occasionally (e.g., monthly), and 5 students reported minimal or one-time use.

[1] Regardless, I obtained the ethics committee approval at TU Delft for conducting this study.

*Table 1. The range and percentage of study fields in the questionnaire response*

| **Study field range** | Count (N=100) |
|---|---|
| Artificial Intelligence & Robotics | 37 |
| Engineering & Design | 15 |
| Medicine & Health Sciences | 18 |
| Philosophy, Ethics & STS | 13 |
| Social Sciences & Management | 8 |
| Mathematics & Physical Sciences | 6 |
| Law & Technology | 3 |

Next, I will cover the general impressions and use cases the students reported, touching on the educational role of LLMs that the students perceive, leaving the suggestions for universities for the discussion section.

*Typical academic use cases*

The 100 responses suggest that most students treat LLMs as multi-purpose academic assistants, helpful for writing, coding, and studying theories and concepts. Table 2 below summarizes the main education use cases the students mentioned in the questionnaire. The most prevalent use pattern (27%) concerns the use of LLMs for drafting text, its improvement, and language refinement, e.g., “making my English sound more professional” or “improve the quality of my English writing, basically like Grammarly on steroids.” A second theme (20%) involves using LLMs for learning and conceptual clarification. Many respondents describe LLMs as an interactive explainer: “I’ve asked ChatGPT to break down complicated subjects to make them more understandable,” or “I use them to discuss articles after reading, as a short preliminary literature search, and to refine research questions.” One respondent identified this back-and-forth interaction as “philosophical/conversational muscle building.” Programming support was nearly as prevalent (17%), especially among engineering and computer science students, typically for use cases as in this response: “Writing short snippets of code (<30 lines), debugging code, and explaining error messages.” The brainstorming function, helping to “get started” on essays or projects, to “sketch problem solutions,” or as a way to overcome writer’s block, was also mentioned frequently (13%).

*Table 2. Main student use cases of LLMs in education per the questionnaire*

| Theme | Description | % of responses (N=100) |
|---|---|---|
| **1. Academic writing, editing, and translation** | Drafting introductions, abstracts, paragraphs, essays, rewriting, paraphrasing, grammar and flow checking, translation, finding synonyms, improving English | 27% |
| **2. Studying and concept comprehension** | Explaining theories, definitions, summarizing or synthesizing readings, extracting insights from PDFs, practicing for exams, generating examples and scenarios to assist comprehension | 20% |
| **3. Programming and technical help** | Coding support, debugging, generating small scripts, explaining error messages, completing code (e.g., Co-pilot use) | 17% |
| **4. Brainstorming** | Generating ideas for essays, projects, research titles, or approaches, inspiration, career advice | 13% |
| **5. Research support** | Identifying relevant literature, outlining methodologies, refining research questions | 8% |
| **6. Information retrieval** | Asking factual questions, getting overviews, cross-checking facts | 8% |
| **7. Project assistance** | Writing emails, preparing presentations, creating visuals, managing projects | 5% |
| **8. Curiosity and experimentation** | Trying out capabilities (e.g., Winograd tests, experiments, image creation), exploring limits of LLMs | 2% |

Overall, students value the explanatory and evaluative capacities of LLMs, not only for interpreting texts or concepts, but also for synthesizing and evaluating information, which the students rely on. Increasingly, they understand LLMs as search engines, "as overpowered Google," directly seeking information and sources through a chat interface, e.g., for "finding seminal academic work on a topic" or "checking research ideas to see if they have already been taken." This quote elaborates well on the shared reasoning behind this (e.g., perceived speed and effectiveness), accompanied by healthy skepticism:

> I have mainly used ChatGPT to get information quicker than searching Google. So, for specific questions that I could not find on Google, I would give the bot a prompt and get my information there. I do, however, cross-check the information given because these bots are not always correct.

In some instances, students even turn to LLMs for career advice and identifying further study programs to consider, a testimony of trust in the counselling capacity with which students identify LLMs . As we have seen in the block quote above, this trust is often not blind, as even in answering the general question on LLM use, many students remarked on the need to double-check and verify machine results (more in Table 3 further). One respondent explained that their trust – or mistrust - was based on an initial experiment: "The first time, I used [LLM] to just see what it can do, for example, asking it difficult questions from the Winograd test. After knowing what it can do, I used it multiple times to come up with [...] plans and [...]

recommendations," to contrast with this response: "I tried to use [LLMs] to rephrase an abstract but discarded the results." I will explore these different ways of using LLMs through the dominant metaphors that the students mentioned in describing LLMs (e.g., An oracle, a tutor), describing in detail the benefits and drawbacks of using LLMs the students identified in the questionnaire.

These findings from 2023-24 are consistent with the latest literature analysis, systematically synthesizing typical LLM use in engineering education, citing brainstorming, explanation of concepts, coding assistance, and overcoming "fear of the blank page" as frequent uses (Akpan et al., 2025; Naznin et al., 2025; Samala et al., 2025; Tillmanns et al., 2025). However, recent literature also suggests that students also start refusing LLMs' use in learning even when the teachers allow this (Dai, 2025; Hanshaw & Sullivan, 2025), citing principled, environmental, ethical, and other reasons. While cases of AI non-use were also reported in the questionnaire, there were only several of them. The user experience and knowledge about LLMs since 2023 could potentially explain this difference, enabling diverse critical appropriation. The typical academic use cases identified in the questionnaire are still relevant and informative insofar as they give a snapshot of what students value in education and the way they position the role of LLMs in that. To uncover this further, I will now turn to the way students used metaphors in this regard.

**Using metaphors to highlight the ethical underbelly of LLMs in engineering education**

As students begin to incorporate LLMs into their academic routines, a set of recurring metaphors emerges that they use to articulate what these tools are, what they are good for, and where their limitations lie. Examining such metaphors is valuable because it reveals how people interpret a phenomenon and the qualities or expectations they project onto it (Black, 2019). In Technology Assessment, metaphor analysis has proven useful for showing how figurative language subtly guides both the design and subsequent use of technologies (Grin & Grunwald, 2000; Grunwald, 2014). Within the philosophy of technology, too, metaphor analysis has become a key lens for understanding how people domesticate new technologies, often before they become widespread in society, offering insight into how users imagine these technologies should function and, conversely, the kinds of users and moral commitments that are presumed within those imaginaries (Sand, 2025; Vallor, 2024; Kudina, 2024).

Metaphors about technology thus play a dual role: they serve as cognitive scaffolding that helps people navigate and interpret unfamiliar tools, and at the same time, they participate in shaping technological development and uptake (Kudina et al., 2025). They are, therefore, not merely descriptive but carry implicit assumptions about how a technology should be used and by whom. In what follows, I examine the two most prevalent metaphors[2] students employed

[2] Other metaphors mentioned referred to a calculator (only once, conveying a sense of technological inevitability, truthfulness, and ease of output) and to "Grammarly on steroids" (several times, commenting on polishing the grammar and style). Because many of the assumptions and concerns in these metaphors recur in the metaphors of oracle and tutor, I focus on these two for reasons of space. For a detailed calculator metaphor analysis regarding LLMs in education, see Guest et al. (2025, pp. 14–15)

in the questionnaire to describe LLMs: an oracle and a tutor. In doing so, I aim to illuminate the hopes, expectations, and anxieties that shape students' encounters with LLMs in educational settings. Table 3 below provides a summary of the main issues students see in this regard, to which I'll be referring when analyzing students' metaphors next.

*Table 3. The problematic aspects in using LLMs for education per the questionnaire*

| Cluster | Description | Illustrative quotes | Count |
|---|---|---|---|
| 1. Quality of output | LLMs often produce incorrect, fabricated, or misleading information with confidence | "It doesn't have a ground truth, it's just bullshitting"; "Makes up sources and facts"; "Terrible with math and physics" | 53 |
| 2. Verification burden | Constant need to (cross)check outputs, adding time, effort, and requiring expertise | "You should always recheck the information"; "Hard to verify if the info is correct"; "Time-consuming" | 34 |
| 3. Ethical concerns | Worries about biased outputs, energy use, privacy, harmful content, manipulation | "Biased or discriminatory answers"; "Privacy concerns"; "Could convincingly spread misinformation" | 19 |
| 4. Impairing the learning process | Fear that LLMs' use weakens learning, reduces creativity, makes students dependent | "Can make the user lazy"; "Not using your own brain"; "Decreases creativity or your own writing voice" | 18 |
| 5. User domain knowledge | Effective use requires crafting precise prompts with context, challenging for many | "I'm scared that I don't have the knowledge to differ good and bad outputs"; "You have to ask the right question"; "Hard to formulate prompts" | 10 |
| 6. Lack of transparency and "ghost sources" | LLMs do not (properly) disclose information sources, provide made-up or unverifiable sources | "It produces URLs to non-existing pages"; "Don't know where data comes from"; "Ghost sources" | 13 |
| 7. Academic integrity and misuse | Generating assignments, giving unfair advantages and undermining academic integrity | "Students can cheat or plagiarize"; "LLMs writing essays for them"; "Fraud" | 12 |
| 8. Generic or repetitive output | Outputs are repetitive, generic, or lack insight, being unsuitable for nuanced work | "Repetitive"; "Monotonous writing style" | 8 |

*LLMs as an oracle*

Students frequently invoke the metaphor of *an oracle* to describe the perceived benefits of LLMs in academic contexts, reflecting an expectation that these systems provide swift and authoritative access to reliable information, as if merging the truth-teller and prophet from Delphi with the digital tools. Many praise LLMs for outperforming conventional search engines in speed and convenience, and for their time- and effort-saving ability to refine vague queries into precise answers. The quote below from one of the respondents exemplifies this:

> LLMs can also (if they can access the internet) improve searching, as they can check out multiple sources for you and use that knowledge (and also the non-internet-y, paper-based resources they were trained on) to answer almost any question you might have very well and also tailor it perfectly to what you were actually asking, as well as explain the answer and give additional details you wouldn’t get from a simple search.

Such associations are shaped by specific interface and interactional design features. The visually clean input field and conversational format construct a sense of frictionless dialogue, reinforcing the impression of interacting with an intelligent interlocutor rather than a statistical model (Narayanan Venkit et al., 2025; Sun et al., 2025). Students also describe LLMs as capable of identifying multiple sources and synthesizing them into neatly tailored explanations, which strengthens their perception of LLMs as an expert truth-teller.

However, this metaphor obscures the underlying technical reality of LLMs as probabilistic text generators rather than fact-retrieval systems (Bender et al., 2021). The moral perception of intelligence and expertise encouraged by the interface design, e.g., clean interface, no visible citations, no encouragement to verify, creates an ethical tension, as it invites users to assume epistemic trust in outputs generated through statistical approximation. Students themselves articulate this friction. While some view LLMs as “very useful tools for searching information,” others point out that the absence of transparent verification mechanisms leads to epistemic vulnerabilities. This worry is further intensified by the emerging evidence of GenAI-produced papers that start populating scientific search systems, such as Google Scholar, which the authors explain as a manipulation of the scientific evidence pool on divisive topics (e.g., climate change) (Haider et al., 2024). The oracle metaphor thus glosses over the gap between user expectations of factual authority and the model’s inability to distinguish truth from plausible-sounding fiction. This mismatch raises concerns about moral or epistemic harm arising from misplaced trust in seemingly intelligent systems.

The oracle metaphor also imposes particular moral expectations on LLM users. While students assume that LLMs function as expert information retrieval systems, they still see the value of critical judgment. The summary of problematic aspects of students’ use of LLMs in education, as presented in Table 3, identifies questioning the quality of output and the need for constant evaluation as the top two concerns. However, the students’ responses also suggest that adopting a “trust-but-verify” position is difficult to sustain in practice, represented by the cluster “User domain knowledge” (Table 3). As one student explains, it “takes skill and practice to learn how to doubt LLMs,” particularly when outputs appear fluent, authoritative, and well-structured. Another student elaborates further: “It is hard to verify if the information is correct. If you ask it for some pointers to papers or websites to get a more in-depth explanation or simply verify it because you want a source to cite for your paper, it produces URLs to non-existent web pages and papers.” Thus, the oracle metaphor amplifies both hopes (e.g., speed, truthfulness, convenience) and worries (e.g., accuracy, verification burden, epistemic over-reliance) about implementing LLMs in learning practices, revealing the deep ambivalence students navigate when pursuing AI-generated information.

*LLMs as a tutor*

Students often frame LLMs as *a tutor*, emphasizing their ability to facilitate deeper academic engagement for explanation, evaluation, learning, and even personal development. This metaphor is strengthened by interface features that simulate pedagogical responsiveness: a conversational flow, continuity across turns, human-like formulations, and affective user-affirming responses. Such technological affordances create a sense of individualized attention, aligned with educational pursuits of personalized learning: “When something is not entirely clear, you can simply ask it to explain that particular section better. This gives a low barrier individual learning experience,” as one student put it. Some students describe asking highly specific, course-level questions and receiving answers that feel attuned to their needs: “It is like asking an expert in the field and you can even tell him how to answer your question.” Yet, LLMs do not track understanding, scaffold misconceptions, or adapt instruction in the pedagogical sense; they generate plausible continuations in a dialogue that *feels* tailored. Moreover, recent studies indicate that the user-flattering and user-affirming emotional language in LLMs enhances the perception of trustworthiness and competence of this technology (Brun et al., 2025; Cohn et al., 2024), entrenching confirmation bias and undermining the learning process by failing to instill rigor (Goetz et al., 2023).

The students also point to these concerns, as one of the most frequently mentioned worries (N=18) related to impairing the learning process by reducing opportunities for reflection and inducing overreliance (see Table 3). Several note that they “learn less” when the system produces outlines or solutions that bypass productive struggle, raising the risk of academic deskilling: as one student put it, “I think they take away from your learning by doing the work for you.” Research on learning suggests that such struggle is not incidental but essential for long-term retention and conceptual understanding (Bjork & Bjork, 2011). The frictionless interface, mirroring patterns already noted in the oracle metaphor, minimizes cues that would encourage users to pause, reflect, or question. As a result, overreliance becomes an easy default.

A subtler but very serious risk arises when students seek help outside their area of competence. A recent study on the use of LLMs by undergraduate mechanical engineering students found that, although popular LLM chatbots, such as ChatGPT, Gemini, and Co-pilot, can provide satisfactory answers to concept-based questions, they overwhelmingly fail at numerical problem-solving and “deep conceptual understanding” (Akolekar et al., 2025, p. 1). The students in my study also report that while LLMs may perform adequately for familiar or well-referenced material, their performance falters in complex theories and STEM-related fields. This student quote echoes a general response sentiment:

> It's terrible with math and also makes mistakes with physics concepts if they get too advanced. It also tends to make up things when you ask it about active areas of research, e.g. in biology. So I would say it's good for base level knowledge, but for specifics it can be erroneous so you definitely shouldn't rely on it 100%.

This trend in the student responses is supported by recent research (Waldo & Boussard, 2024), suggesting that the further students move from areas where they can independently judge correctness and evaluate the coherence and tone of information, the harder it becomes to notice when LLMs present incorrect or misleading information, or express hidden biases.

This is significant for learning overall and the tutor metaphor specifically because bias in LLMs is a well-documented issue. Multiple studies analyzed bias in LLMs, e.g., in perpetuating stereotypes and social biases (Gallegos et al., 2024), diminishing the visibility of women's perspectives in shared knowledge systems (Barry & Stephenson, 2025), exacerbating disparities in underrepresented groups (Mishra, 2025), or promoting dangerous health and safety advice (Oviedo-Trespalacios et al., 2023). A recent bias study is particularly fitting for analyzing the tutor metaphor, as it shows how ChatGPT adapts the length, language, and style of explaining STEM theories by inferring the learner's gender from the language formulations (Hedlund, 2025). Thus, in the explanation of scientific concepts, females are more often presented with narrative or domestic metaphors, while males - with technical, process-focused explanations. My study suggests that even though the students list general ethical concerns, including bias, as the third most-referenced problematic category in using LLMs (N=19, Table 3), they do not outline specific biases or the manner in which they materialize in LLMs' use. For both students and teachers, this communicates an urgent need not only to be aware of the presence of such biases but also to have the ability to spot and mitigate them to maximize the learning process.

However, as discussed in the oracle metaphor, this is precisely the area where students struggle the most. As one student mentioned (original spelling preserved):

> It can be hard to distinguish when it [LLM] hallucinates. Typically using it for clarification on topic you have already covered in a course can be very usefull [*sic*]. But using it to learn something new or to come up with something on the edges of knowledge can yield wildly wrong results which can be dificult [*sic*] to spot when not equipped with the propper [*sic*] basic knowledge about a topic.

Students often describe feeling uncertain about verifying the content and logic of the LLM's explanations, capturing the asymmetry of their own expertise vs the perceived expertise of LLMs. The tutor metaphor, therefore, risks overstating the reliability of LLMs and wrongly imbues them with pedagogical intentionality, encouraging trust precisely where the students' ability to evaluate accuracy or bias is weakest. In this sense, the metaphor shapes expectations that exceed what LLMs can support pedagogically and epistemically.

*A cruel optimism of LLMs*

Before I proceed to the overall concluding discussion for the chapter, I wanted to briefly reflect on the analysis in this section. When asked about their overall impressions of using

LLMs for education, amid a mixture of generally positive, negative, and unsure remarks, *the predominant trend (42%) was providing a positive evaluation, with a caveat*. The group of students here acknowledges the possible utility of LLMs and evaluates them positively, praising their speed, efficiency, and user-friendliness, while at the same time mentioning the work LLMs necessitate from the students to make these technologies effective, e.g., the work of verifying sources, soundness of arguments, double-checking the information provided, and overall vigilance. An exemplary response succinctly represents this duality, suggesting that LLMs are "a great tool if you use it in the right way and know what to be careful of." The students' optimism is, thus, not naïve but informed by experience and a critical lens. At the same time, this quote, one of many similar responses, suggests that instead of placing blind trust in the LLMs and their output, it is the students' responsibility to both be aware of this technology's possibilities, limitations, and ethical considerations, while at the same time, having the domain expertise and the academic skills to be able to verify the credibility and soundness of the output. This response tendency concurs with the frequent educators' practice of allowing the LLMs' use, emphasizing that it is ultimately the students' responsibility for the output. Given the ethical analysis of the metaphors above, I wonder whether this expectation is realistic, considering that students, even in the higher education context, are still there to learn, both in terms of knowledge production and evaluation.

Consistent with this trend of positive LLM evaluation, albeit with a caveat, the responses also *convey the overall desirability of their introduction in higher education*, provided that proper AI literacy skills are in place. The quote below gives a good example of the layered and value-laden logic in this case:

> I think LLMS are pretty neat. They are easy to use and give helpful information for someone to use. However, I also think it could be risky. We don't know what is going on in the machine and we don't know if the given answers are correct or completely made up. From experience I found that after a while the LLM starts to make things up instead of actually giving the right information. This is why I think we should embrace the LLMS and educate people on how to use them and what the limits and risks of these models are.

This quote nicely captures the overall analysis of the metaphors and the questionnaire: the promises of LLMs' utility and efficiency are so strong in the respondents' eyes that the students are willing to forgive them all the practical and emotional work they put into making these technologies effective, as well as the ethical risks that the students very reflectively outlined in their use of LLMs. The metaphor's analysis also exemplifies a predominantly affective relation to this technology as embodying, e.g., the pursuit of personalized and on-demand education, the ideal of learning efficiently, the added value of being an early adopter, while also pursuing the search for truth and harnessing the skills of effective knowledge communication.

In general, *the prevailing metaphors in the students' responses suggest a sense of "cruel optimism,*" drawing on the work of Berlant (2020). This cruel optimism arises "when something you desire is actually an obstacle to your flourishing" (Ibid, p. 1). The LLM-related metaphors convey the values of efficiency and truth (i.e., an oracle), clarity and correctness (e.g., a tutor) that students perceive in LLMs. Abstracting from them allows identifying the desire for frictionless and fast learning through instant answers and methodological shortcuts. Students want LLMs to work in order to study more deeply and quickly, to handle the overwhelming study workload, and to gain (communicative) confidence. These aspirations transform LLMs not just into a functional tool but, in Berlant’s terms, into an object of attachment, associated with desires for one’s own flourishing in learning.

Herein lies the main contradiction, as exemplified in the quote above, the “cruelty” of the attachment to the umbrella of LLM’s promises. While LLMs are portrayed as an oracle or a tutor, they routinely produce misleading, biased, and fabricated output that requires constant vigilance from the students and manual verification. Therefore, the perceived or desired efficiency collides with the actual required vigilance and output evaluation. Here, the promises of ease of use and time-saving do not match the additional, and at times significant, cognitive, emotional, and moral labor required to proofread LLMs’ outputs and align them with the original expectations. *For this evaluation to be effective, the students must apply the skills and knowledge they may not yet possess, which is precisely why they are pursuing higher education* (e.g., disciplinary knowledge, source identification and validation, epistemic humility, etc.). This transforms LLMs into a constant maintenance practice that depends on the very expertise students are seeking to gain at the university (while this maintenance may, in a broad sense, be a part of the educational trajectory and still be pedagogically valuable – but not in the sense of students’ metaphors). Pointing to the epistemic irresponsibility of LLMs and the mounting students’ responsibility in this regard, Flenady and Sparrow (2025) produce a similar evaluation: “[T]he expectation on tertiary students to assume responsibility for their so-called ‘tutors’ and ‘collaborators’ is pedagogically perverse, amounting to a demand that students take sole responsibility for the accuracy of claims they are not able to properly assess” (p. 1).

The students are quite aware of the technological limitations, which they were keen to mention throughout the questionnaire, even when not prompted. Nonetheless, as with other technological innovations that don’t quite deliver on their potential (e.g., voice assistants and the hope of “natural” interaction (Kudina, 2021)), the seductive technological promises that often speak to the core of individual practical and moral aspirations often keep the students returning to using LLMs, forgiving their disappointments and failures. Still, the cruel optimism of LLMs is only self-defeating insofar as people continue to believe in the promised value of LLMs’ effectiveness and efficiency while performing the bulk of the overviewing and evaluative work themselves. Therefore, harnessing any value of LLMs in education requires shifting perspectives on what they are and, crucially, what they are not by design (e.g., tutors or oracles), critically reappropriating their limitations into pedagogical opportunities and more often than not, choosing alternative educational tools.

**Moving forward: pragmatically, cautiously, principally – but always purposefully**

Working through the students' responses in the questionnaire, accompanied by the recent literature review of LLMs' introduction in education, allows me now to shift to a concluding discussion and switch perspectives from students to individual educators and universities. When prompted with the last question in the study, "What should higher education institutions do about LLMs, if anything at all?," an overwhelming majority of responses urges academic institutions to teach (both teachers and students) about the responsible and ethical use of AI in learning (N=41) and to provide clear rules and guidelines (N=27), as illustrated by this quote: "Help students use LLMs optimally and communicate clear rules about when and how to use them." Even though the students want clear rules and guidelines, *it is not always possible and desirable to have black-and-white solutions*. Accounting for the ethical risks, the university should be a space to foster curiosity and experiment in a controlled manner – staying connected with the students' needs and practices, not pushing them into the underground, but acknowledging them and building a dialogue with them. That's why, assuming a general educators' awareness of the technological opportunities and limitations, in this chapter, I propose a purpose-driven and context-based approach.

*Any potential adoption of AI in education should be purpose-driven*. The purpose can be clarified through answering the questions, such as "Why are GenAI applications (or other AI technologies) needed in this course/assignment?," "What added value, if any, do they represent?," and "How exactly would this contribute to the learning objectives?" While this may start at the level of aspiration and curiosity, controlled experimentation with AI in class should be assessed for evidence vis-à-vis the initial motivation. If the evidence does not support original intentions or does not substantially contribute to them, it should be a clear indicator to object to AI's use for the specified purpose. Because, for every potential use, the educator must position the potential pedagogical benefits against the known risks (even as listed in Table 3).

One of such particular risks, mentioned only once in the student questionnaire, is the *environmental harm of GenAI use*. Even though by now there is growing societal awareness about the large resource intensiveness of developing (Gen)AI systems (O'Brien, 2024; Li et al., 2025; Seessel, 2023; Strubell et al., 2019; Lacoste et al., 2019), the individual use of GenAI systems, e.g., chatting with LLMs, usually flies under the radar because of the seemingly negligible individual costs in energy or water. For instance, Google suggests that its LLM application Gemini uses as little energy per prompt as powering a microwave for one second (i.e., 0.24 watt-hours) and as little water as 5 drops (Crownhart, 2025). However, the recent UNESCO report on the sustainability of AI use (2025) puts things in perspective, zooming in on just one popular LLM application, ChatGPT:

> As of June 2025, ChatGPT receives approximately 1 billion queries daily, each using around 0.34 Wh of electricity, about what it takes to power a high-efficiency LED lightbulb for a few minutes [...].That adds up to roughly 310 GWh per year, which is comparable to the annual electricity consumption of over 3 million people in Ethiopia (p. 5).

Additionally, energy-intensive AI development and daily operations in data centers require a lot of water to cool this infrastructure: “AI demand is expected to consume between 4.2 and 6.6 billion cubic meters of water by 2027, surpassing Denmark’s total annual water withdrawal” (Ibid., p. 6). Considering the constant increase in different LLM applications, their increasing individual daily use, and default embedding in educational digital infrastructures (e.g., Microsoft’s Co-pilot), *the energy and water costs of GenAI use* are far from trivial and *need to be taken into account as a disvalue to sustainability next to potential pedagogical value*. I’m not suggesting that individual educators make a detailed cost-benefit analysis for each task, but I do urge to keep these high environmental costs in mind and include them as an overall consideration in deciding whether and how to use (Gen)AI in the classroom. In cases of doubt or no clearly supported purpose, it’s better to err on the side of caution and not use GenAI because of the known downsides.

If the pedagogical utility is identified, it is best “*incorporating AI as a supplementary tool, guiding students toward critical thinking rather than passive reliance*” (Akolekar et al., 2025, p. 9). For instance, using LLMs to generate questions or answers and having the students evaluate them is particularly relevant for engineering and computer science fields with numerical emphasis and coding practice. An example from my philosophy courses in the engineering curriculum is a learning-by-doing exercise on the LLM debate. Here, the students are first asked to prepare and stage an in-class debate about evaluating AI’s use in socially-sensitive contexts based on pre-read materials (e.g., on algorithmic assessment of citizens), and then are asked to replicate the same debate conditions in LLMs, evaluating their answers from the domain expertise they obtained in the first part of class and from the perspective of academic skills (e.g., proper referencing, logical lapses, argumentative strength, etc). In this way, AI is present both as an object of evaluation and through a controlled use, stimulating critical thinking and collective deliberation (Ceres, 2023; Kudina & Muravyov, 2025). Another approach would be to target epistemic injustices in knowledge ecosystems by prompting LLMs to intentionally diversify the syllabus or the references LLMs suggest (Kudina et al., 2025). These examples go in line with the critical pedagogical suggestions of Efimova and Nygren (2025): “If generative AI is used to represent public discourse for learning purposes, it should be scaffolded with a focus on missing perspectives, corroboration, and contextualization and conceptualization of AI-generated content” (p. 10).

A different example of using AI in class in an augmented manner is from the mathematics field, where researchers (Portegies et al., 2025) developed, tested, and validated an educational AI tool to help students develop the skills of writing mathematical proofs. Called “Waterproof,” it combines more traditional rule-based AI (i.e., relying on predefined logic rules to structure data and make decisions) that provides a library of proof examples and invites the students to develop their own proofs in a declarative manner, explaining the results and logic. The GenAI addition provides a chat-based interface that helps with students’ challenges in a fluent, interactive way. The combination of rule-based and generative AI components mitigates the risk of misinformation (GenAI’s “hallucinations”) and induces the students not only to develop the proofs themselves in a creative manner but also to reflect on the process. This example is consistent with the suggestions that, particularly in engineering, AI is best used to support the metacognitive skills of students, geared at creative and evaluative thinking (Feng et al., 2025), in a way that accompanies teacher-led learning (Simelane & Kittur, 2025).

In addition to reviewing learning methods and activities, it is essential to adapt or redesign assessment methods to account for the possibility of intentional or incidental AI use when this was not originally planned. On top of that, the students raise a worry of potential inequality in learning and assessment when the use of AI goes unchecked: "As a student I grapple with the concept that I may be competing for grades, not with other students, but with AI." Another response hints at the systemic inequalities that accompany AI use beyond the grades: "Some individuals might use them to write entire assignments for them and thus not do the work as assigned. Some individuals might not have access to these tools or not understand how to use them and thus another who is using them has some advantage." The earlier mention of "unchecked" does not mean relying on the GenAI detection software because it has so far proven to be inaccurate and unreliable, as well as risking instilling the policing dynamic in learning (Ardito, 2025). Rather, it refers to the reflective adaptation of the assessment methods that account for AI's possible (mis)use. This call also emerges from the student responses (N=27), who are eager to suggest how this assessment redesign may look like: "Let students write essays in person so they still learn the skill," "Shift grading from memorization to analysis," "Use oral exams along with paper exams to ensure quality," or "Make students present their work or answer personalized questions about it," to name just a few. These are all relevant suggestions that, overall, point to a need to go beyond the reproducibility of knowledge and to adjust the assessment process, possibly to closed methods, digital or analogue.

Next to these ad-hoc suggestions, broader assessment frameworks that could correspond with different degrees of depth of possible AI use in class are emerging. One noteworthy AI-aware assessment framework is AAA (Against, Avoid, Adopt) by Lye and Lim (2024). Here, if AI can complete the lower-order skills per Bloom's taxonomy (e.g., recalling facts, explaining concepts or theories), the authors recommend keeping the assessment for these tasks supervised (e.g., ad-hoc Q&A sessions, debates, oral assessment, closed digital or hand-written exams) (*Against*). Next, for the higher-order skills (e.g., analyzing, evaluating, creating), where AI currently performs with mixed results but that students may nonetheless be tempted to use, Lye and Lim suggest relying on contextualization in current affairs, personal experiences, in-class events, and integrating human interaction when possible (*Avoid*). Finally, for the small subset of tasks and in the spirit of a learning-by-doing approach, the authors suggest designing some assessments that use AI intentionally (*Adopt*), letting students brainstorm with AI or critique its output while still submitting their own work and reflecting on their AI use. This could, for instance, mean letting the students complete a coding assignment with the GenAI tool, while asking them to submit a personal evaluation and a de-bugging report, which emphasizes critical thinking and self-reflection (Feng et al., 2025).

Whichever way the education curriculum embeds GenAI, *preserving an element of technological liminality, of uncertainty that goes with LLMs*, is useful here. In engineering ethics, rephrasing the Collingridge dilemma (1980), this refers to the in-between stage of technological design and implementation: after its initial design, when the direction of the consequences is difficult to anticipate and influence, but before the technology is fully entrenched in society and any consequences that go with it are hard to change due to the systemic embedding (Mertens, 2018). Today, we are witnessing the active back-and-forth between the students and teachers and contestation by society at large of the promises of tech companies about the purpose and usefulness of LLMs mirrored in increasing academic and mass media publications, public debates, and governmental policies. This individual and collective effort characterizes precisely such a technological in-between space, a fruitful space for maintaining technological flexibility to some extent and collectively influencing technological trajectory in society. Within universities, cultivating such technological ambivalence could mean preserving the space for alternatives and opting out of AI use (Shata, 2025), providing freedom to contest GenAI tools and discuss the surrounding anxieties (Verano-Tacoronte et al., 2025), and maintaining pedagogical flexibility with an eye to different tools, rather than going all-in on (Gen)AI. To maintain educational resilience, preserving a spirit of learning and curiosity, and being open to dialogue remain essential, with GenAI being only the latest case in point. As Friesen (2020) put it,

> From the printing press to personalized learning, new pedagogies and technologies, each in their time, have been configured in remarkably similar ways in educational discourse: they are seen as overcoming political compromises, human failings, even the 'dark' ways of the past; and they are regarded as ushering in a kind of pedagogical utopia of natural, authentic, even playful teaching and learning. This in turn gives the present a sense of urgency. It, in turn, is portrayed as a time when action, investment and change—often unprecedented in scope and scale—are all urgently needed (pp. 141-142).

With this historical word of caution on the implementation of technology in teaching and learning practices, it will not only be intellectually curious but also pedagogically important to reevaluate this chapter in the future. Because what is at stake is reinterpreting and reasserting the value of education as Generative AI matures and as we, educators and students, go through several hermeneutic cycles, (re)inventing our own ways to fit with - and challenge - this disruptive technology.

## Further reading

**Notes on contributor**

Olya Kudina is an Associate Professor in Ethics/Philosophy of Technology at TU Delft, exploring the interaction between values and technologies, often through the lens of empirical philosophy. Her recent monograph is “Moral Hermeneutics and Technology” (Rowman & Littlefield, 2024).